\documentclass[10pt, sigconf]{acmart}

\usepackage[english]{babel}
\usepackage{blindtext}
\usepackage{eso-pic}
\usepackage{todonotes}
\usepackage{titlesec}
\usepackage{listings}

\usepackage{hyperref}
\usepackage{multirow}

\graphicspath{{figure/}{figures/}}

\renewcommand\footnotetextcopyrightpermission[1]{}
\setcopyright{none}
\usepackage{eso-pic,xcolor,enumitem}

\usepackage{color,soul}
\usepackage{xspace}

\usepackage{ifthen}

\newboolean{showcomments}
\setboolean{showcomments}{true}
\ifthenelse{\boolean{showcomments}}
{ \newcommand{\mynote}[3]{
    \protect\fbox{\sffamily\scriptsize#1}
    {\small$\blacktriangleright$\textsf{\emph{\color{#3}{#2}}}$\blacktriangleleft$}}}
{ \newcommand{\mynote}[3]{}}

\newcommand{\ty}[1]{\mynote{Tianyuan}{#1}{teal}}

\usepackage{subcaption}

\usepackage{etoolbox}
\usepackage{xstring}
\DeclareListParser{\doslashlist}{/}

\newcounter{ndnNameComponentCounter}%
\newcommand{\name}[1]{{%
  \setcounter{ndnNameComponentCounter}{0}%
  \renewcommand{\do}[1]{{%
    \ifnumgreater{\value{ndnNameComponentCounter}}{0}{\allowbreak/}{}%
    \ifnumodd{\value{ndnNameComponentCounter}}{}{}%
    \detokenize{##1}}%
    \stepcounter{ndnNameComponentCounter}}%
``{\fontfamily{cmtt}\selectfont\IfBeginWith{#1}{/}{/}{}\doslashlist{#1}}''%
}}

\usepackage{amssymb}
\usepackage{url}
\usepackage{tabularx}
\usepackage{xurl}

\usepackage[algoruled,vlined,noend,linesnumbered]{algorithm2e} 
\usepackage{spverbatim}

\SetAlFnt{\small}
\SetAlCapFnt{\small}
\SetAlCapNameFnt{\small}
\SetAlCapHSkip{0pt}
\IncMargin{-\parindent}
\setcopyright{none}

\begin{document}
\title{Hydra - A Federated Data Repository over NDN}

\author{\textbf{Justin Presley, Xi Wang, Tym Brandel, Susmit Shannigrahi} \\\textit{Tennessee Tech}}

\author{\textbf{Xusheng Ai, F. Alex Feltus} \\\textit{Clemson University}}
\vspace{1em}
\author{\textbf{Proyash Podder,  Alex Afanasyev} \\\textit{Florida International University}}
\vspace{1em}
\author{\textbf{Tianyuan Yu, Varun Patil, Lixia Zhang} \\\textit{UCLA}}
\date{\today}

\begin{abstract}
Today's big data science communities manage their data publication and replication at the application layer. These communities utilize myriad mechanisms to publish, discover, and retrieve datasets - the result is an ecosystem of either centralized, or otherwise a collection of ad-hoc data repositories. Publishing datasets to centralized repositories can be process-intensive, and those repositories do not accept all datasets. The ad-hoc repositories are difficult to find and utilize due to differences in data names, metadata standards, and access methods. To address these problems, some communities use high-level frameworks such as iRODS. However, since these solutions depend on location-oriented TCP/IP protocols, they need to hide the complexity of creating location-independent services over TCP/IP, making them tightly coupled and complex to build and maintain. While storing and accessing all data from commercial cloud platforms could be another solution, such platforms are costly for storing and serving large amounts of data.

To address the problem of scientific data publication and storage, we have designed \textit{Hydra}, a secure, distributed, and decentralized data repository made of a loose federation of storage servers (nodes) provided by user communities. Hydra runs over Named Data Networking (NDN) and utilizes the State Vector Sync (SVS) protocol that lets individual nodes maintain a ``global view" of the system. Hydra provides a scalable and resilient data retrieval service, with data distribution scalability achieved via NDN's built-in data anycast and in-network caching and resiliency against individual server failures through automated failure detection and maintaining a specific degree of replication. Hydra utilizes "Favor", a locally calculated numerical value to decide which nodes will replicate a file.
Finally, Hydra utilizes data-centric security for data publication and node authentication. Hydra uses a Network Operation Center (NOC) to bootstrap trust in Hydra nodes and data publishers. The NOC distributes user and node certificates and performs the proof-of-possession challenges.

This technical report serves as the reference for Hydra. It outlines the design decisions, the rationale behind them, the functional modules, and the protocol specifications. 
\end{abstract}
\maketitle
\pagestyle{plain}

\section{Introduction} \label{sec:introduction}
This tech report describes Hydra, a distributed and decentralized data repository system built over NDN. The motivation for this work comes from the needs of scientific communities such as genomics, climate science, and high-energy particle physics that often follow a distributed collaborative model where data is published and accessed by scientists worldwide. Hydra provides an easy-to-use platform for publishing and accessing such datasets through a distributed, federated storage system where a specific community or a project provides individual storage nodes./ 

This tech report gathers all the Hydra design decisions and the lessons learned through the design process in one place. The report also describes the implementation of Hydra, how it provides users with secure and scalable file publishing and sharing services built using the NDN protocol stack, and our experience of Hydra's first deployment on the FABRIC testbed\cite{8972790}.

Hydra is designed to run over a federation of storage servers provided by different user organizations. Users can publish files to Hydra, which can be shared securely and scalably following defined access policies. Hydra also maintains a consistent "system state" among all the storage servers utilizing the NDN State Vector Sync (SVS) protocol\cite{8599772}.
Hydra automatically replicates files to maintain a desired degree of replication even in the face of individual server failures.

Specifically, Hydra is designed to achieve the following goals.
\begin{enumerate}
\item Reducing scientific communities' data publication and management burden. Currently, data is published through ad hoc systems or centralized repositories, none of which work well with the rapidly exploding distributed datasets. Hydra provides a decentralized framework that allows scientists to publish datasets easily and scalably.
\item Enabling fast and reliable data replication across distributed nodes. Once a node fails, Hydra automatically replicates files to the available nodes, and the system can continue to run after the failure without operator intervention. 
\item Improving data repositories' performance and supporting more users and applications by distributing read and write requests to different data nodes.
\item Improving data findability and reusability of data by addressing data by their names. This approach will allow the communities to spend less time curating data and tracking their locations once they agree on a project-specific namespace. 
\item Incorporating data-centric security in the design. All data in Hydra are signed and publicly verifiable.
\end{enumerate}

Hydra uses the data-sharing model in the genomics community as the primary use case. In this community, data is generated in a distributed manner by various research facilities. The researchers then send the data to central facilities (e.g., repositories hosted by National Center for Biotechnology Information or NCBI) that make them available publicly. Apart from the centralized model, this approach has two problems that impede efficient data sharing. First, data publication through a central repository includes a manual component (checking and processing for quality and conformity to naming standards) and can take a long time (weeks or months) before data is available. Second, these central repositories accept specific types of datasets. Datasets that are not accepted by central repositories have to be published in ad hoc ways, making them very difficult to find and utilize in research. Hydra makes this process automated and simpler through the expressive naming of content. Once datasets are named according to community-accepted standards, they can be immediately published through Hydra and discovered and utilized by the users and workflows.

\section{The Design of Hydra} \label{sec:design}

\subsection{Design Goals} 
\label{subsec:design-goals}
We aim to build a federated, distributed data repository system over NDN, where individual storage servers are contributed by the users in the community and are geographically distributed. Here are the design goals of Hydra:

\begin{itemize}

\item All Hydra operations should use a data-centric security model. Our design secures Hydra as a distributed system and can secure the files stored in Hydra. To ensure the security of the Hydra federation, each Hydra server goes through security bootstrapping process to obtain its security credentials before being deployed from a Network Operation Center (NOC). All the files stored in Hydra are cryptographically signed for authenticity and can be encrypted for confidentiality. 


\item Hydra should enable nodes to operate under different administrative domains. A Hydra federation should be able to operate despite differences in the amount of resources and policies.

\item Hydra should support file insertions with automatic replication and scalable file retrieval. 

\item The Hydra federation should perform semi-autonomously. Operators can perform infrequent operations such as configuring trust and managing catastrophic failures manually, but normal operations such as nodes joining the federation, recovering from node failure, and replication of data should not need user involvement or operator intervention.


\item Other desired performance measures for Hydra include the ability to support (a) publications of large datasets (GB to TB in size),  and (b) disk-to-disk throughput of 1Gbps or more.

\end{itemize}

\subsection{Design Assumptions}
\begin{itemize}
\item The number of nodes in a Hydra system is expected to be tens of nodes in the initial deployment; we hope to gain further insights through experimentation on how well Hydra can scale in terms of the number of nodes.

\item In its initial trial deployment, Hydra does not hold the master copy of the data. Instead, it used the storage as a scratch space for publishing short-term data. All data that has not been used in a month will be automatically deleted. This assumption will be re-visited as Hydra becomes mature.
    
\item As a collection of user-contributed storage servers, we assume some nodes may malfunction (that lead to file errors and node unavailability), but there is no malicious node in the federation.

\item Each User file insertion request is not an interactive step, which is made infeasible by the potentially long delay in carrying out a file insertion.\footnote{The size of files could be multiple gigabytes, thus uploading files to Hydra may take multiple minutes.}
Instead, the notification regarding success/failure will be provided to the user via a status URI. 
    
\item Hydra has a central identity manager (NOC). NOC is where the trust originates. Hydra defines trust schemes based on names and previously defined permissions. Hydra uses Google OAuth to establish identities and enforce trust schemas. Nodes are pre-authenticated. They use a bundled certificate distributed with the code.

\item In Hydra, we assume that the NOC is trusted, and each Hydra node is also trusted after it completes the bootstrapping process.
\end{itemize}

\subsection{Design Approaches} \label{subsec:design-approach}
Here are the design approaches for Hydra:

\begin{itemize}

\item For all new and existing data under our control, we adopt the principle in naming. For all existing data using URLs as unique names, we gradually remove the ``location" semantic in those names, treating them simply as unique data identifiers.
\item All user interfaces will be kept simple. 
\item Operations such as file insertions, deletions, and fetching will be based on the file names.
\item The users will not be exposed to the internal workings of Hydra, such as file replication and location information of the storage servers. 
This is not needed in Hydra since NDN routing allows users to reach the ``best" node. 

\item We take on a decentralized approach in Hydra design, even though there may be an extra cost of decentralization - communication overhead.

\end{itemize}

\section{Hydra Functional Details} \label{sec:overview}

Hydra is a federated storage system where participants from a big data community (e.g., genomics) contributes storage. Hydra utilizes NDN to create a federation of storage servers. The Hydra federation allows users to publish data into the system using data names agreed-upon by the community, making them readily available and lowering the barrier of data publication.


\begin{figure*}[!ht]
    \centering
    \includegraphics[width=\textwidth]{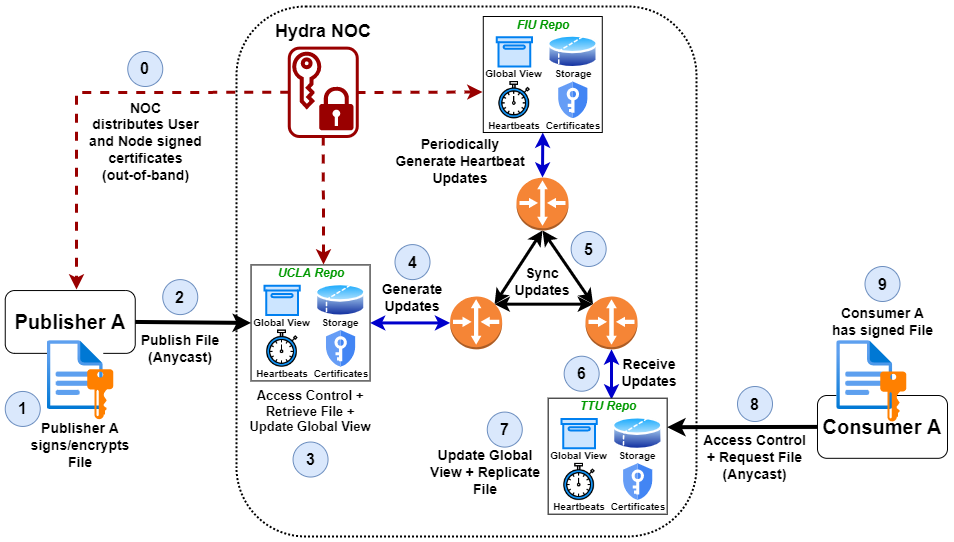}
    \caption{A High-level overview of Hydra}
    \label{fig:overview}
\end{figure*}

 Figure~\ref{fig:overview} outlines Hydra's general structure along with an overview of the functions performed by Hydra. In its simplest form, a Hydra federation has several geographically distributed nodes. 
 The Hydra federation also relies on a NOC that is not part of the federation but distributes certificates to the nodes and publishers. 

The process of inclusion of a node into Hydra begins by creating a default route to one of the Hydra nodes and requesting a certificate from the NOC. The NOC authenticates the requests and returns a certificate that the node uses for signing further communication. This completes node bootstrapping. Once the security parameters are bootstrapped a node goes through system boostrapping where it creates its own global view. 
The command is routed to a Hydra node (decided by NDN routing) that verifies the command and ingests the file. Once the file is ingested, a group message is triggered. Other nodes subscribing 
receive the notification, update their global view, and some of the nodes start to replicate the file. Each successful replication publishes a group message that allows the nodes to keep track of the degree of replication. The nodes also send out periodic heartbeats to the other nodes so that failure recovery can begin when a node goes offline. Finally, a user or a workflow retrieves a file by sending an Interest with the name of the file. This Interest is routed to Hydra nodes that can return the file.

\subsection{Function Overview} \label{subsec:functions}

\begin{figure*}[!ht]
    \centering
    \includegraphics[width=2\columnwidth]{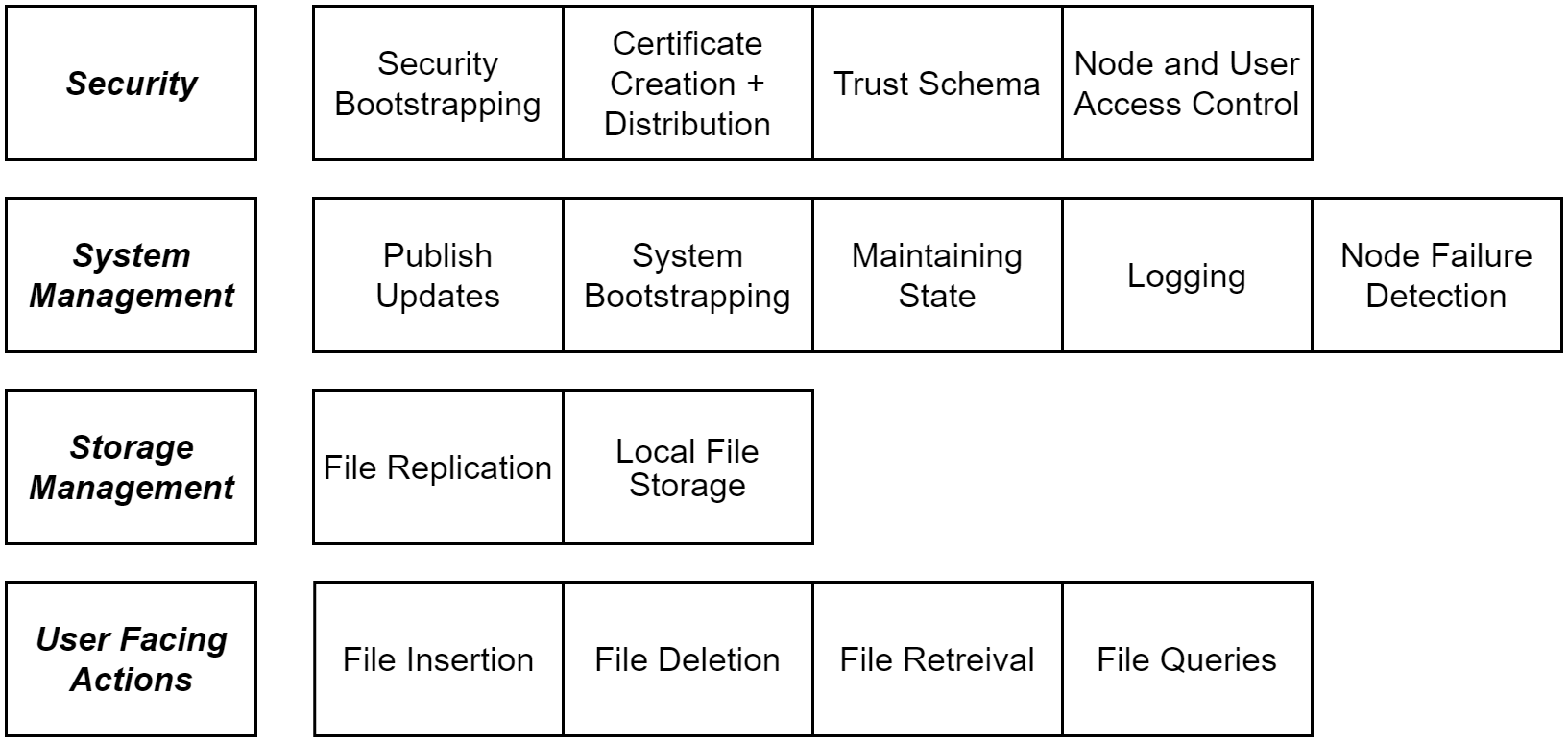}
    \caption{Necessary Functions of a Hydra Node}
    \label{fig:node-functions}
\end{figure*}

As Figure~\ref{fig:overview} outlines, a Hydra federation has four types of functions:

\begin{itemize}
    \item Bootstrapping
    \item System Management
    \item Storage Management
    \item User Facing Functions
\end{itemize}

Figure~\ref{fig:node-functions} outlines each of these categories.

\subsubsection{Bootstrapping} can be divided into security bootstrapping and system bootstrapping. Security bootstrapping focuses on getting proper key and certificate information that the node can present to the federation before it is allowed to join. We discuss bootstrapping in Section   \ref{sec:node-operations}. System bootstrapping focuses on the functionality needed for a new node to join the federation and sync the node's state to the system's state. 

\subsubsection{System Management} modules provide the necessary systems level functionality for Hydra. These include maintaining states in the network, publishing and applying updates to the system state, 
logging, node failure detection, and failure recovery. These functions are at the core of how Hydra operates. Management of system is discussed further in Section \ref{sec:node-operations}.

\subsubsection{Storage Management} functions oversee file replication and local file storage. The file replication module replicates files across Hydra nodes and maintains the degree of replication (currently 3). This ensures data loss does not occur even when individual nodes fail. The local file storage module handles the storage functions of a node. These include interacting with the underlying database or filesystem, and freeing up storage by deleting unused files.
Further discussion on file replication can be found in Section \ref{sec:data-operations} and local storage can be found Section \ref{subsec:modules}.

\subsubsection{User Facing functions} provide user interfaces for various user and application interactions. These functions include file insertion, deletion, retrieval, and system status checks. Hydra exposes name-based APIs for these functions that can be directly used by the users or integrated into separate applications. More details on user facing functions can be found in Section \ref{sec:data-operations}.
\subsection{Modules} \label{subsec:modules}
\begin{figure}[!ht]
    \centering
    \includegraphics[width=\columnwidth]{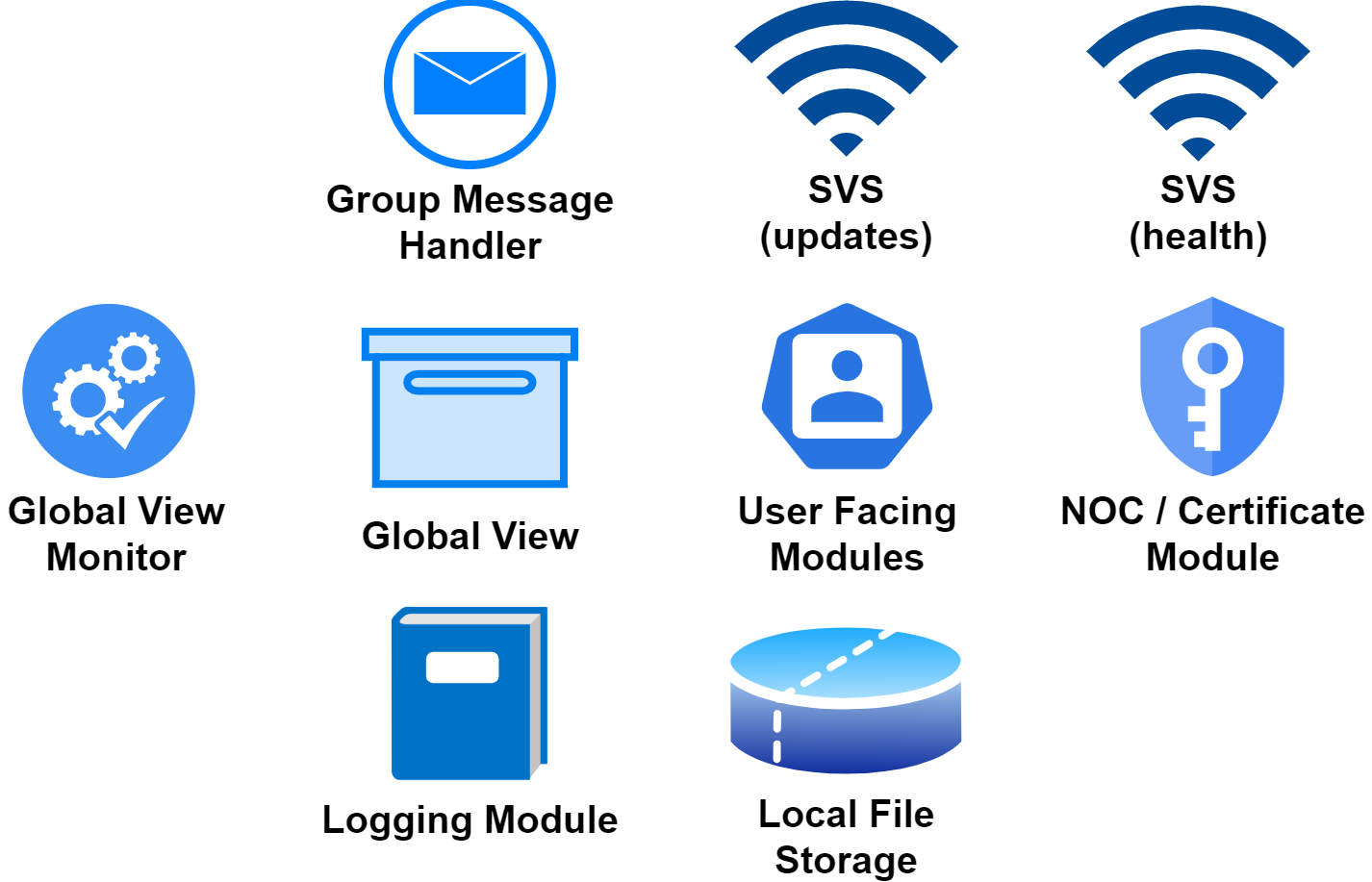}
    \caption{Modules that make up a Hydra Node}
    \label{fig:node-modules}
\end{figure}

To provide the functions described in Subsection~\ref{subsec:functions}, a Hydra node is made up of several modules as illustrated in  Figure~\ref{fig:node-modules}. Nine modules make up a Hydra node, and the following list briefly describes each of them. 



\subsubsection{StateVectorSync (SVS)}

The SVS module uses the SVS protocol to publish data (e.g., state update messages) that is received by all other nodes within Hydra. SVS allows Hydra to maintain a consistent view across the nodes.

\subsubsection{Global View}
The Global View is a local database to a Hydra node. It represents a node's view of the Hydra federation. The global view contains information related to the Hydra system such as files' metadata, node information, and replication information.

It also contains the heartbeat tracker sub-module that is used to send and track a heartbeat message. The heartbeat message establishes whether a node is alive or unavailable.  If the nodes in the federation do not hear a heartbeat message from a node for a certain time, it presumes the node is unreachable and sets the node in the heartbeat tracker to ``Unreachable". Once $n$ heartbeats are missed from a specific node, the nodes initiate automatic replication.


\subsubsection{Group Message Handler}

All Hydra nodes join a ``group" via SVS and synchronize the global view by publishing group messages. Group messages posted by a node are visible to all nodes. This module handles Group Messages - messages that are sent to the group of Hydra nodes. The group message handler interacts with SVS to receive and apply the new updates to Global view and publish new group messages.

\subsubsection{Global View Monitor}
The global view monitor monitors the global view database for any changes. If a change is detected, it initiates an outgoing update message. It also applies update messages received by the node to the Global view database.

\subsubsection{User-Facing Modules}

These public-facing modules allow users to interact with the Hydra system, providing data insertion, deletion, and retrieval functionality. They also allow the user/applications to query the system for file availability. 

\subsubsection{Network Operation Center (NOC)}
NOC is a centralized entity that acts at the root of trust for a deployment. NOC provides signed certificates to the nodes and users. 

\subsubsection{Logging Module}

This module logs all node records (data operations and errors) to monitor Hydra and provide information to the administrator.

\subsubsection{Local File Storage}
Local file storage is the Database on a node that holds the Data packets corresponding to a published file. The file storage module also performs garbage collection and deletes any file not accessed for a certain duration.

\section{Hydra Design Specifications}\label{sec:specification}
This section outlines the design specifications of Hydra.
\subsection{Naming} \label{sec:naming}
Names in Hydra are important as they are the primary construct of the system.

Hydra uses several types of names:
\begin{itemize}
    \item The Hydra Prefix: decided by the operator(s).\\ Example: \name{/Hydra/}
    \item File (content) Names: decided by data publisher(s). \\Example: \name{/human/genome/dna/hg38}
    \item Node Names: decided by the operator(s).\\ Example: \name{/hydra/node-X1}
    \item The NOC Prefix: decided by operator(s).\\ Example: \name{/hydra/NOC}.
\end{itemize}



\subsubsection{Hydra Prefix}
The <hydra-prefix> acts as the primary namespace for Hydra operations. An example Hydra prefix can be \name{/hydra} or \name{/genomics}. 
This hydra prefix utilizes SVS to actually distribute messages sent on this prefix to all nodes in the federation. 

Below are a few more specific uses of sub-namespaces under the Hydra prefix.
\begin{itemize}
    \item \name{/<hydra-prefix>/<function>/<function-info>} is utilized to conduct functions that are user-facing. Users use this prefix for inserting, deleting, and any other interaction that affects data directly within the network. An example might be \name{/hydra/insert/<filename>}.

    \item \name{/<hydra-prefix>/group} is the prefix for group communications. This prefix is utilized for sending messages through State Vector Sync between nodes. The group prefix is under the Hydra prefix since it is used to notify all the nodes of changes in the system.
    
\end{itemize}

\subsubsection{Node prefixes}
The node prefix is utilized for unicast operations which allow Interests to be sent between a specific node and a specific target node within Hydra. The main use case of the node prefix is to enable communication between Hydra nodes. The node prefix is also used for directing a user request to a specific node if the contacted node does not have the content. It is also used for replication.
The node prefix is in the form of \name{/<node-name>/<hydra-prefix>/fetch/...}. This can be appended by operation or content specific names.

\subsubsection{ForwardingHints}
When data is not present in the node that a user contacted, it returns the name of the node that holds the data. This forwarding hint allows the user to express a subsequent Interest that is steered to a specific node name. 
One example might be \name{/human/genome/dna/hg38} with a forwarding hint to \name{hydra/node-X1}.


\subsubsection{Name Prefixes for NOC}

NOC is considered the trust anchor of Hydra. While it is not part of the Hydra federation, it uses NDN names for receiving requests. The name prefixes for NOC would be \name{/hydra/NOC/...}.



\subsubsection{Node Naming}


Nodes names are required to be unique. Hydra does not add any extra components to a node name when using it for communication. 
Example names can be
\name{'/TnTech/CSC/hydra'} or \name{'/hydra/node-xx'}, or \name{'/norm-labs/hydra-5'}.

\subsubsection{Certificate names}
Hydra uses a hierarchy of certificate names to provide trust. 

An example Hydra root key (hydra root = hydra trust anchor) would be \name{/hydra/KEY/…}.

Node cert names can be \name{/hydra/32=nodes/host.ucla/KEY/… (key for UCLA node)} or 
\name {/hydra/32=nodes/fabric-hostXYZ/KEY/… (key for hostXYZ)}.

The user cert names can be  \\
\name{/hydra/32=users/XX@gmail.com/32=ns/FIU/experiments/KEY/...} or 
\\
\name{/hydra/32=users/YY@gmail.com/32=ns/FIU/experiments/2022/KEY/...}

\subsubsection{Content naming}
There are two ways we can name files in Hydra. First, we can utilize the content names created by the publisher. Example of such a name would be \name{/human/genome/dna/hg38}. The other option is to utilize Hydra specific names such as \name{/hydra/human/genome/dna/hg38}. While both are acceptable, there are a few trade-offs. With the first option, we assume the owner of the name prefix (i.e., \name{/human/genome} will allow Hydra to announce the prefix. Second, if publishers publish content using different name prefixes, a Hydra node must announce all those name prefixes into the routing system (e.g. \name{/human/genome} and \name{/kidney}). Finally, creating an easy-to-utilize trust schema  might be difficult with this model. 
On the other hand, appending the \name{/Hydra} prefix to each name makes trust relations and routing announcements simpler. A node can simply announce the \name{/Hydra} prefix into the routing system. However, this requires changing the content names (e.g./ \name{/human/genome/dna/hg38} becomes \name{/hydra/human/genome/dna/hg38}, which makes the content specific to the Hydra framework. For our implementation, we use the first naming model.

\subsection{Security} \label{sec:security}

A Network Operation Center (NOC) acts as the system trust anchor for Hydra deployments, and issues security certificates. A new Hydra node fetches a certificate by sending an Interest. Next the node verifies the returned self-signed certificate out-of-band. Then the node signs the Interest and requests a certificate from the NOC. Finally, NOC signs the certificate using the trust anchor and returns it in the replied data.

In order to authenticate the Hydra group communications and interactions with users, Hydra’s security model includes three roles in the system.
\begin{itemize}
\item Network Operator Center (NOC): serving as the system trust anchor for each Hydra deployment, adding or deleting Hydra nodes from the federation, and managing the security policies for the Hydra system.
\item Hydra Node: a server that joins the federated repo system. One trusted server might start a new process if the old one terminates.
\item Client: users who use the Hydra system by sending Insertion and Deletion commands or data requests.
\end{itemize}

In Hydra’s security design, we assume the NOC is trusted, and each Hydra node is also trusted once added. To bootstrap one Hydra node to the system, out-of-band verification (e.g., pre-shared passcode, email verification) is needed to authenticate the initial communication between the Hydra node and NOC. A new Hydra node fetches the trust anchor by sending Interest /hydra/bootstrap/anchor, and verifying the returned self-signed certificate out-of-band. Then it uses the out-of-band pre-shared material to sign Interest /hydra/bootstrap/cert and request a certificate from the NOC. Verifying the certificate requester’s authenticity, the NOC uses the trust anchor to sign a certificate and returns it in the replied Data. The new Hydra node learns its node name from the certificate via certificate installation. After installing the trust anchor and node certificate, the Hydra node keeps its security policies up-to-date by periodically send Interest /hydra/bootstrap/schema and fetch new policies.
\\The Hydra group communication should also be protected from invalid message senders. Therefore, Hydra requires each node to sign the Interest and Data in group communication and enforce the system membership checking through verifying the signer’s certificate with the trust anchor.
\\Besides that, users’ commands to the Hydra system should also be authenticated. Different from authenticating the group communication among Hydra nodes, Hydra users do not share the trust anchor with the Hydra nodes. The identity of users can only be verified along the certificate chain when the Hydra nodes have users’ trust anchors. Hydra achieves this via security policy distribution, where individual Hydra nodes fetch the trusted anchors and corresponding signing rules under those namespaces.
\subsection{Files} \label{sec:files}
Hydra uses the term \emph{file} to describe the data unit of insertion, deletion, retrieval, or replication. However, a file is not necessarily a file in the UNIX system; it is just a BLOB of data that is identified by a unique name. See Section~\ref{sec:naming} for more details on file names.

A file consists of NDN Data packet(s); however, the size of a file directly impacts the performance of the following operations: replication, failure recovery, insertion, and retrieval. These NDN Data packet(s) also determine the unit size of a particular file.


 \subsection{Favor} \label{sec:favor}
Due to Hydra being a federation of storage nodes with different storage capacities and policies, it is necessary to have a mechanism that can express a node's local conditions and preferences for storing or replicating additional files.

\textit{Favor} can be thought of as "how suitable a node is to carry file(s)".
Every node is responsible for calculating a numeric value for node X where X is each node within the system including itself. The range of this numeric value is determined by the formula used. As implicitly indicated, favor is an entirely local calculation. However, all parameters of the calculation are announced by every node giving each node complete control over what and which files are stored in their respective storage. Essentially, favor protects node independence within the federated system.

As it stands, favor is calculated per node and all parameters are included in every GM in order to be as up to date as possible. To limit network traffic, Hydra does not send a GM when these parameters change. See Section~\ref{sec:group-messages} for more details on GMs.

Favor calculations may incorporate the following aspects:
\begin{itemize}
    \item storage capacity (current usage and max usage)
    \item network stability / traffic
    \item location
    \item prefix preferences, file origin, node's labels, etc
    \item local policy
\end{itemize}

Favor affects Hydra's overall performance, usability, and stability. For our initial implementation, we simplified favor's calculation to be based only on storage capacity.

In the future, additional parameters and functions of favor may be necessary for different applications. For example, a baseline favor value may be necessary for high-volume storage environments to reserve some capacity for file takeovers essentially protecting Hydra's resiliency. See Section~\ref{sec:future-work} for more details on future work. \subsection{Storage} \label{sec:storage}
A Hydra node utilizes several databases for maintaining state and for storage of different types of data.

Each of these databases is described briefly below.
\begin{itemize}
    \item SVS Database: A database of all published messages over the Sync group, implemented as a key-value store.
    \item Global View: A Hydra node's state -- a relational database containing all information relating to the entire system.
    \item Local File Storage: A key-value database holding files that the Hydra node is storing.
    \item Local Reserved Storage: A sectioned off part of Local File Storage to only be used to help with special operations such as data ingestion. This space is not used for storing replicated files or Hydra's metadata.
    \item Command Table: A table holding the progress and status of the commands being currently executed.
    \item Logs: These may either be stored locally or inside a distributed logging system or both.
    \item Certificate and Key Database: A key-value database that holds key and certificate information required for secured publication of messages, data validation, and user authentication.
\end{itemize}

Currently, we do not impose any limit on these storages. However, operational deployments will need to define the storage limit for these databases.
  file insertions, deletions or failures to all other nodes. Hydra uses the State Vector Sync (SVS) protocol\cite{svs} to achieve efficient and loss resilient global view synchronization.
SVS is further discussed in Subsection~\ref{sec:svs}.

While GMs help exchange states, we still need a way to transfer and synchronize GMs between hosts. For this purpose, Hydra utilizes the State Vector Sync (SVS) protocol. 
Note that this sync protocol only provides eventual consistency guarantees. Note that SVS only provides a synchronization capability for message exchange. Hydra is able to utilize any other synchronization protocol for message exchange.


  \subsection{Group Messages} \label{sec:group-messages}
Hydra utilizes Group Messages (GMs) to exchange updates among nodes.
A published Group Message goes out to every node within Hydra such that every node receives every Group Message. 
Understanding  all past actions in Hydra allows the nodes to act in a more precise and knowledgeable manner.

There are six Group Message types with each performing different actions. 
\begin{itemize}
    \item Insert: a GM containing metadata of a new file that is inserted in Hydra.
    \item Delete: a GM containing information about a file to delete from Hydra.
    \item Claim: a GM stating the node is going to fetch or is unable to fetch a certain file.
    \item Store: a GM stating a node has stored a file in its local storage. 
    \item Heartbeat: a GM that is periodically published by nodes to let others know that it is alive. In our current implementation, the interval of heartbeats is set to 30 seconds. 
    \item Leave: a GM send by a leaving node in order to help speed up replication
\end{itemize}

To keep network overhead low, we treat all messages as heartbeat messages. If there is no GM withing a specified time, a specific heartbeat message is announced.
\subsection{Global View} \label{sec:global-view}
The Global View can be thought of as 
a node's view of the entire system.

As nodes exchange messages (Group Messages, see next section), a Hydra node creates and stores the state of the system in a light-weight local database known as the 'Global View'. 


Note that this Global View is not stored once in a global storage accessible by every node. Instead, each node has its own version of this global view that they maintain. Throughout the duration of operation, a Hydra node synchronizes this database with the other nodes through continuously published group messages.


Every node in Hydra has a global view of:
\begin{itemize}
    \item All node information
    \item Each file’s specifics
    \begin{itemize}
        \item which nodes are in possession of the file (the “on list”)
        \item which nodes can step up to take over the file (the “backup list”)
        \item meta-info (size, origin node, copies, and etc)
    \end{itemize}
\end{itemize}

The Global View also has the ‘state vector’ that made up the Global View. This state vector is the same type found in SVS and indicates the sequence of messages that were incorporated into the state.

Global View layout example (capital alphabetical letters are placements for node names):
{\small
\begin{spverbatim}{
"state_vector": {"A":123,"B":120,"C":100,"D":150,"E":123},

"nodes": [
    {"name": "D", "favor": 50, “alive”: True},
    {"name": "C", "favor": 20, “alive”: True},
    {"name": "B", "favor": 15, “alive”: True},
    {"name": "A", "favor": 10, “alive”: True},
    ...
    ],

"files":[
    {
        "name": "/genomics/fileA",
        "size": 128,
        "contact": "B",
        "copies": 3,
        "on": ["B","A","D"],
        "on_history": ["C"]
    },
    ...
    ]
}\end{spverbatim}
}
\subsection{PubSub Protocol} \label{sec:pubsub}

Hydra's interface for data insertion and data deletion takes inspiration from the previous incarnation of standalone NDN repo\cite{}. A PubSub-like protocol is used for exchanging messages with clients such as file insertion commands and updates. These operations are further outlined in Section~\ref{sec:data-operations}. 
\subsection{Status Codes} \label{sec:status-codes}

Users can receive multiple status codes (like a server error + redirection for instance), the following is an overview of all status codes in Hydra. 

\begin{table}[]
\caption{Hydra Command Status Codes}
\begin{tabular}{|cc|}
\hline
\multicolumn{1}{|c|}{\textbf{Status}}   & \textbf{Code} \\ \hline
\multicolumn{2}{|c|}{\textbf{Informational}}            \\ \hline
\multicolumn{1}{|c|}{STAND\_BY}         & 100           \\ \hline
\multicolumn{1}{|c|}{FETCHING}          & 101           \\ \hline
\multicolumn{2}{|c|}{\textbf{Success}}                  \\ \hline
\multicolumn{1}{|c|}{OK}                & 200           \\ \hline
\multicolumn{2}{|c|}{\textbf{Redirection}}              \\ \hline
\multicolumn{2}{|c|}{\textbf{Client Error}}             \\ \hline
\multicolumn{1}{|c|}{BAD\_NAME}         & 400           \\ \hline
\multicolumn{1}{|c|}{BAD\_REQUEST}      & 401           \\ \hline
\multicolumn{1}{|c|}{UNAUTHENTICATED}   & 402           \\ \hline
\multicolumn{1}{|c|}{UNAUTHORIZED}      & 403           \\ \hline
\multicolumn{1}{|c|}{NOT\_FOUND}        & 404           \\ \hline
\multicolumn{1}{|c|}{NO\_COMMAND}       & 405           \\ \hline
\multicolumn{2}{|c|}{\textbf{Server Error}}             \\ \hline
\multicolumn{1}{|c|}{RESOURCE\_LIMIT}   & 500           \\ \hline
\multicolumn{1}{|c|}{TRAFFIC\_OVERLOAD} & 501           \\ \hline
\multicolumn{1}{|c|}{NODE\_DISCONNECT}  & 502           \\ \hline
\multicolumn{1}{|c|}{UNKNOWN\_ERROR}    & 503           \\ \hline
\end{tabular}
\label{tab:status-codes}
\end{table}

\section{Hydra User Operations}
\subsection{User Bootstrapping}
\label{sec:dataop-bootstrap}
In order to bootstrap a Hydra user into the system, each Hydra user needs to obtain trust anchor, trust policies of Hydra and and get certified by the Hydra NOC

In Hydra, each user obtains the Hydra trust anchor and initial trust policies out-of-band.
After that, Hydra NOC authenticates each user on their email addresses.
This requires Hydra NOC knowing trustworthy email addresses via initial out-of-band trust relations.
We rely on the human trust relations between users and the NOC operator to realize the user authentication. 
The NOC operator maintains a list of trustworthy email addresses and configures the it to the NOC application.

Hydra user utilizes NDNCERT~\cite{10.1145/3125719.3132090} to perform the email authentication and obtains certificate from NOC.
Specifically, Hydra NOC as the NDNCERT CA does the following steps upon receiving a certificate request from a Hydra user.
It first checks the Hydra user email address membership, then verifies the email address by sending a PIN and requesting sending back in NDNCERT message.
Upon successful email membership and possession verification, Hydra NOC uses the user naming convention\ty{where's the user naming convention? Susmit - also what do we use this name for?} to assign the user an NDN name based on its email and certifies it.

As the final step, a Hydra user completes its security bootstrapping by obtaining and installing the issued certificate.

\section{Hydra Node Operations} \label{sec:node-operations}

Aside from handling incoming data operations, a Hydra node has other responsibilities partly due to being in a distributed, federated system and partly due to managing its own resources.

\subsection{Node Security Bootstrapping}
\label{sec:nodeop-boostrap}
Similar to the user bootstrapping (Section~\ref{sec:dataop-bootstrap}), each Hydra node need to obtain its trust anchor, trust policies and certificate.

In Hydra, each node obtains the trust anchor and initial trust policies out-of-band.
After that, Hydra NOC authenticates and authorizes each node on their node names and public keys.
This requires Hydra NOC knowing trustworthy <node name, public keys> bindings via initial out-of-band trust relations (e.g., ssh, email).

For simplicity, Hydra uses email to establish the initial trust relations.
Before establishing a new node, the node operator emails the <node name, public key> binding to the NOC operator who configures the NOC application to authenticate the received binding.

Hydra node utilizes NDNCERT to perform the public key authentication and obtains certificate from NOC.
The Hydra NOC first verifies if the hydra node has the membership in the system by checking its public key in the pre-configured trusted bindings, and it asks the node to perform the Proof-of-Possession Challenge where the node uses private key to sign a nonce to prove its public key possession.
Upon successful signature verification, Hydra NOC issues the new node with a certificate with the node name and public key obtained from trusted name-key bindings.

A Hydra node completes its security bootstrapping by obtaining and installing the issued certificate.

\subsection{Federated Node Responsibilities}

\begin{figure}[!ht]
    \centering
    \includegraphics[width=\columnwidth]{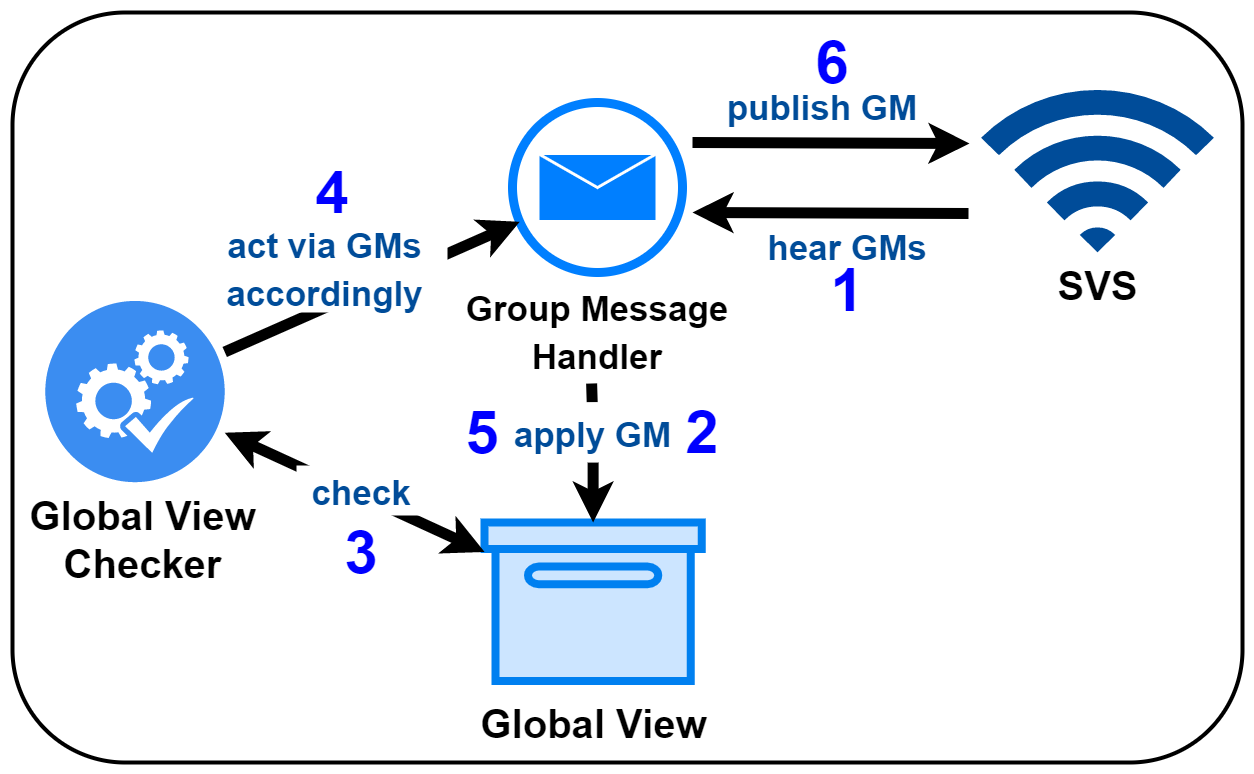}
    \caption{Modules Involved with Checking the Global View}
    \label{fig:checker-sys}
\end{figure}

One of the major responsibilities that a node actively fulfills is checking if any federated actions are required to be taken. This is done by the Global View Checker. The Global View Checker consistently checks the Global View as seen in figure \ref{fig:checker-sys}.

\subsection{Storage Management}
Each node storage space contains two components: Local File Storage and Local Reserved Storage. Local File Storage provides sufficient space to support valid replication requests. The function of Local Reserved Storage is to reserve computational resources for system processes, this space is not used to store replication files or metadata for Hydra.

\subsection{Joining the federation and membership management}

To join the federation a node must install hydra and begin broadcasting a heartbeat. Once the broadcasted heartbeat has been detected between the new node and one of the pre-existing nodes within the federation. The pre-existing node will update the global view for the federation to acknowledge the new node. At the same time, the new node is still broadcasting its heartbeat to other pre-existing nodes within the federation, and they are following suit with the previously discussed pre-existing node's actions. This will increase the propagation of the new node's acknowledgment within the global view. Once this is complete it will be listed as a node within the network or federation, and the others can begin to interact with the new node.
The NOC serves as the centralized node management system within the hydra federation. Node management is inherently done through limiting the behavior of valid node operations to the operations discussed within this section, providing pooled storage, and updates through heartbeat and global view messages to the wider federation. If these node requirements are found to be insufficient for any given node then it is determined by the NOC, which manages adding and deleting nodes from the federation. The root admin with access to the NOC is responsible for the final approval and deletion of a node into the federation.

\subsection{Nodes becoming unresponsive} \label{sec:nodes-unresponsive}
\begin{figure}[!ht]
    \centering
    \includegraphics[width=\columnwidth]{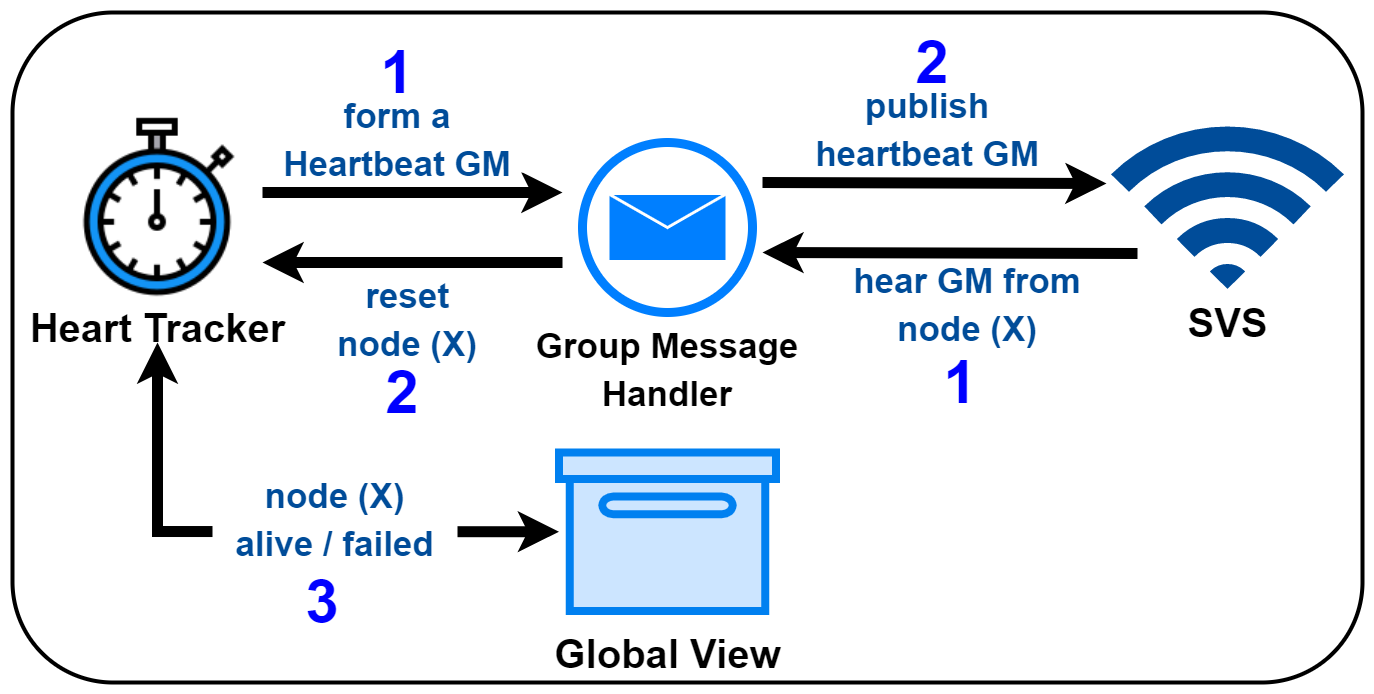}
    \caption{Modules Involved with Tracking and Maintaining Node States}
    \label{fig:heartbeats-sys}
\end{figure}

A node becomes unresponsive for the following reasons:

\begin{itemize}
    \item To leave the system a node must remove the hydra infrastructure or shut down. It will become unresponsive and be dropped from the system over time providing the same availability to rejoin that is discussed in 6.5.
    \item An incident occurs that limits communication between two or more nodes, or completely shut down the node.
\end{itemize}

To label a node as unresponsive, a total of three failed heartbeats must be noticed by another nodes heartbeat tracker. The Heartbeats system including the tracker is shown in figure \ref{fig:heartbeats-sys}. All nodes are responsible for their own detection of unresponsive nodes. Once an unresponsive node has been detected, the investigating node will move the unresponsive node to a suspended record of nodes that have gone offline. This status allows the node to update its record for other operations such as managing heartbeats and file replication. No message through State Vector Sync (SVS) or the Global View updating the status of the unresponsive node is sent out over the network by the investigating node.

This is done for two reasons:
\begin{itemize}
    \item First, it allows unresponsive nodes during some disruptions to continue their communication with the rest of the network in the case of a partition. It is also assumed that all other nodes will recognize that a node is unresponsive too if it had truly gone offline. Once the disruption is over the unresponsive node will follow the reestablishment process with the investigating node(s) that have labeled it unresponsive.

    \item Second, if a node has completely gone offline; it can be assumed all other nodes will soon detect this because the offline node will no longer send heartbeats.
\end{itemize}

These two simple assumptions for handling unresponsive nodes cover all situations that could be reached within the system and allow for more accurate detection of disruptions within the system by handling them locally. 
This is done to avoid situations where node A may communicate normally with the rest of the network except for node B. If we did not handle unresponsiveness locally in this case, node B may alert the rest of the network that node A is unresponsive, cutting it off from communicating with the entire network due only to a minor communication problem between node A and B.

\subsection{Nodes leaving the system}
Nodes at anytime may leave the system as they wish and must simply state it as any other update To leave the system, a node can publish the leave GM or become unresponsive as previously discussed in section \ref{sec:nodes-unresponsive}. The leave GM allows the rest of the system to begin the replication procedure outlined within section \ref{sec:nodes-unresponsive}. This simplifies determining what data needs to be replicated as it can be determined from the leaving node itself, rather the the federation of nodes determining which files. Once the leaving node has determined which files need to be replicated it will publish the leave GM containing this list for the other nodes. Once a leave GM is published the other federation nodes can replicate this list of files received. However, a leaving node is not required to stay for this process, since the federation can recover from it becoming unresponsive. A leaving node deciding to remain throughout the leaving procedure does provide performance benefits to replication as replicas can be obtained from the leaving node.The leaving node will begin replicating files to them based on the favor found within all GMs.

In this formal leaving process it is not necessary for the other nodes to need to determine what files to replicate. They will allow the leaving node to determine this has it already has the knowledge of how many copies of its own files exist, and which need to be replicated. Other federation nodes are on stand by until the leaving node becomes unresponsive or completes the formal leaving process. Once this process is determined complete from store GMs a second leave message will originate from the leaving node to alert the federation of this, and it will complete leaving the system. During this stand by all other federation nodes will move the leaving nodes on list records to the 'on history' records. The leaving node will be listed in the 'on history' for a duration of one month in case the leaving node rejoins the federation. When this month duration expires all records of the left node will be removed from the 'on history'. This allows for temporarily leaving the system but if a node is gone for to long treating it as a new node upon rejoining.


\subsection{Reestablishing a Node's State / Node State Reinstatement}
For reinstatement of a node the same name of the node's previous state must be used. This is done to re-link with the records other nodes within the network hold for the previous iteration of our current node. Along with this a local recovery file representing the state in which said node was at the time it became unresponsive or offline is utilized. This file is used locally to determine any differences that have occurred on the system since it last communicated with the Hydra network, and then communicate these differences to the wider network. The purpose of using the differences in system states is that it is more likely by only correcting what is different few corrections will be needed overall, this assumption does dwindle overtime. This process is to allow the node to reclaim its previous status within the network while also updating any nodes that need to know about changes. For example, if a data set was removed during the lapse in communication then other nodes expecting our reinstated node to contain that data need to be alerted to the fact it no longer contains that data. This allows for quick and accurate recovery of a node even if files were deleted or inserted from its system during the lapse in communication.

\section{Hydra Data Operations} \label{sec:data-operations}

To demonstrate how Hydra performs data operations, we will utilize an initial deployment on FABRIC.
We assume all these operations happen after the data publisher is authenticated by the Hydra node.


Imagine the scenario where Clemson's Genomics and Bioinformatics facility decides to use Hydra to store pre-processed (i.e. indexed) genomes. The main goal of this exercise is to allow data access for the researchers and their collaborators in a reliable manner.

A user can download the correct set of files and insert into downstream analytic workflows simply by asking for these datasets by name. 


\subsection{Data Insertion} \label{sec:data-insert}

\subsubsection{Scenario}
Alice, a graduate student who is doing research at the Clemson's Genomics and Bioinformatics facility, has produced a pre-processed (i.e. indexed) axolotl genome that she believes is critical in understanding an axolotl's ability to regenerate tissue. She desires to publish this data in the form of a File within Hydra. To satisfy this scenario, she performs the following interactions with Hydra.

\subsubsection{User-to-Node Interaction}
\begin{figure}[!ht]
    \centering
    \includegraphics[width=\columnwidth]{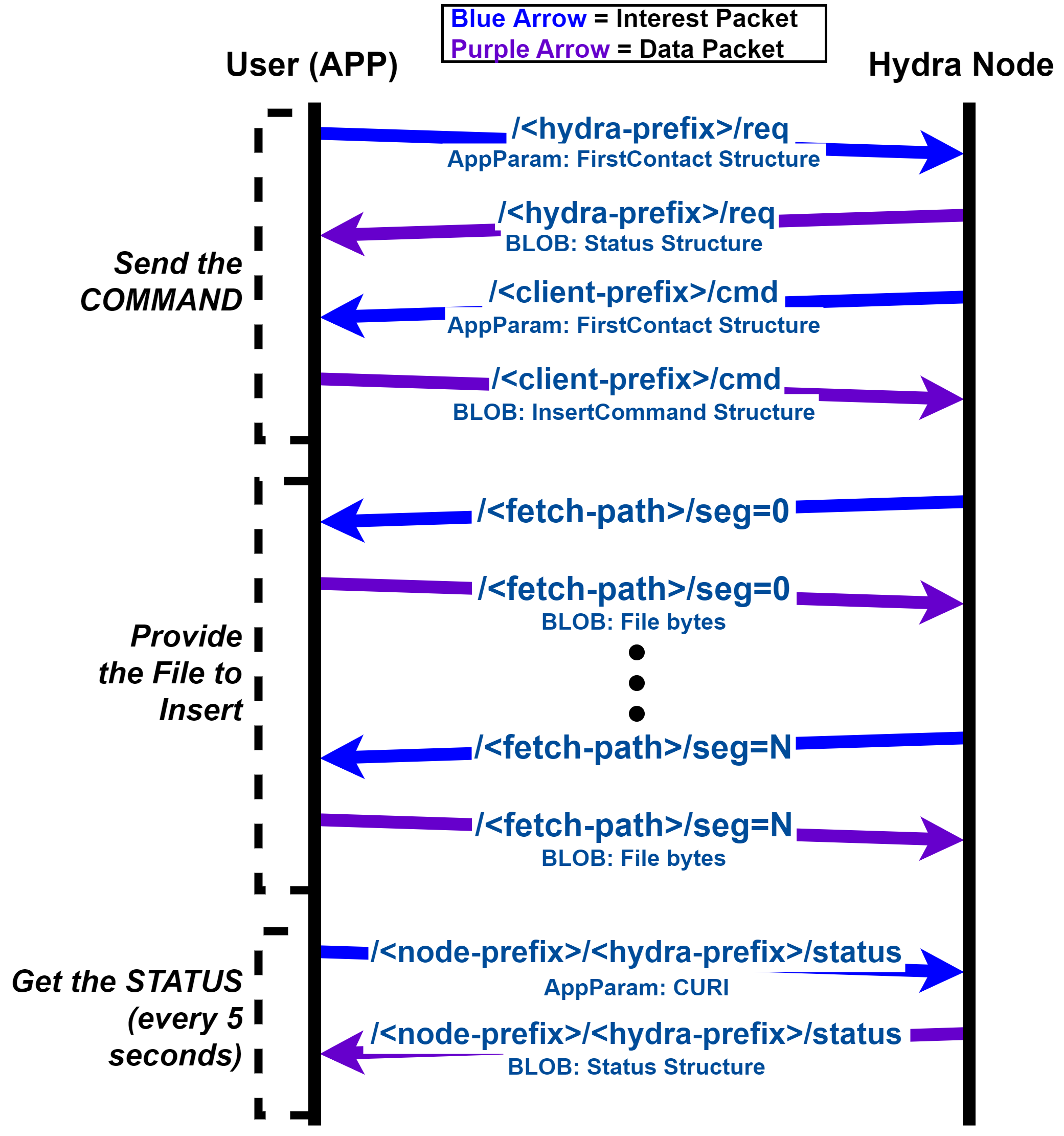}
    \caption{NDN Interaction of A User and A Hydra Node During Data Insertion}
    \label{fig:insert-usr}
\end{figure}

The process of how a user will interact with a Hydra node via NDN is described in Figure~\ref{fig:insert-usr}. Any Hydra node can be contacted by the user for data insertion, and this interaction can be summarized as a PubSub-like interaction: it uses a notification Interest with a component /notify and data retrieval via /msg.
This interaction between user (A) and node (X) goes as follows:
\begin{enumerate}
    \item User (A) will send an Interest notification using the prefix \name{/<hydra-prefix/insert}, announcing that it has a command for any Hydra node to process.
    \item Upon hearing this, node (X) will send an Interest to fetch this command using user (A)'s prefix (stated in user (A)'s FirstContact structure found in the Interest notification's app parameters).
    \item Node (X) will begin to process this command; and because the command is an insertion, node (X) will begin to fetch file (F) using user (A)'s fetch path (found in the InsertCommand structure).
    \item Once file (F) is retrieved, node (X) will send a notification Interest stating that the command that user (a) sent has a updated status.
    \item User (A) can then fetch the status of the command using node (X)'s prefix (stated in node (X)'s FirstContact structure found in a previous interest).
\end{enumerate}

There are a few key points throughout this process.
\begin{enumerate}
    \item User (A) can fetch the status at any time as it has the necessary info; however if the command is not ready, node (X) will tell user (A) to wait. 
    \item The switch from anycast to unicast is necessary to ensure a proper response as command information is not directly shared between Hydra nodes.
    \item Throughout the interaction, a command unique resource identifier (curi) is used. This allows Hydra nodes to process more than one command from a user and allows users to send more than one command.
\end{enumerate}

\subsubsection{Module Interaction}
\begin{figure}[!ht]
    \centering
    \includegraphics[width=\columnwidth]{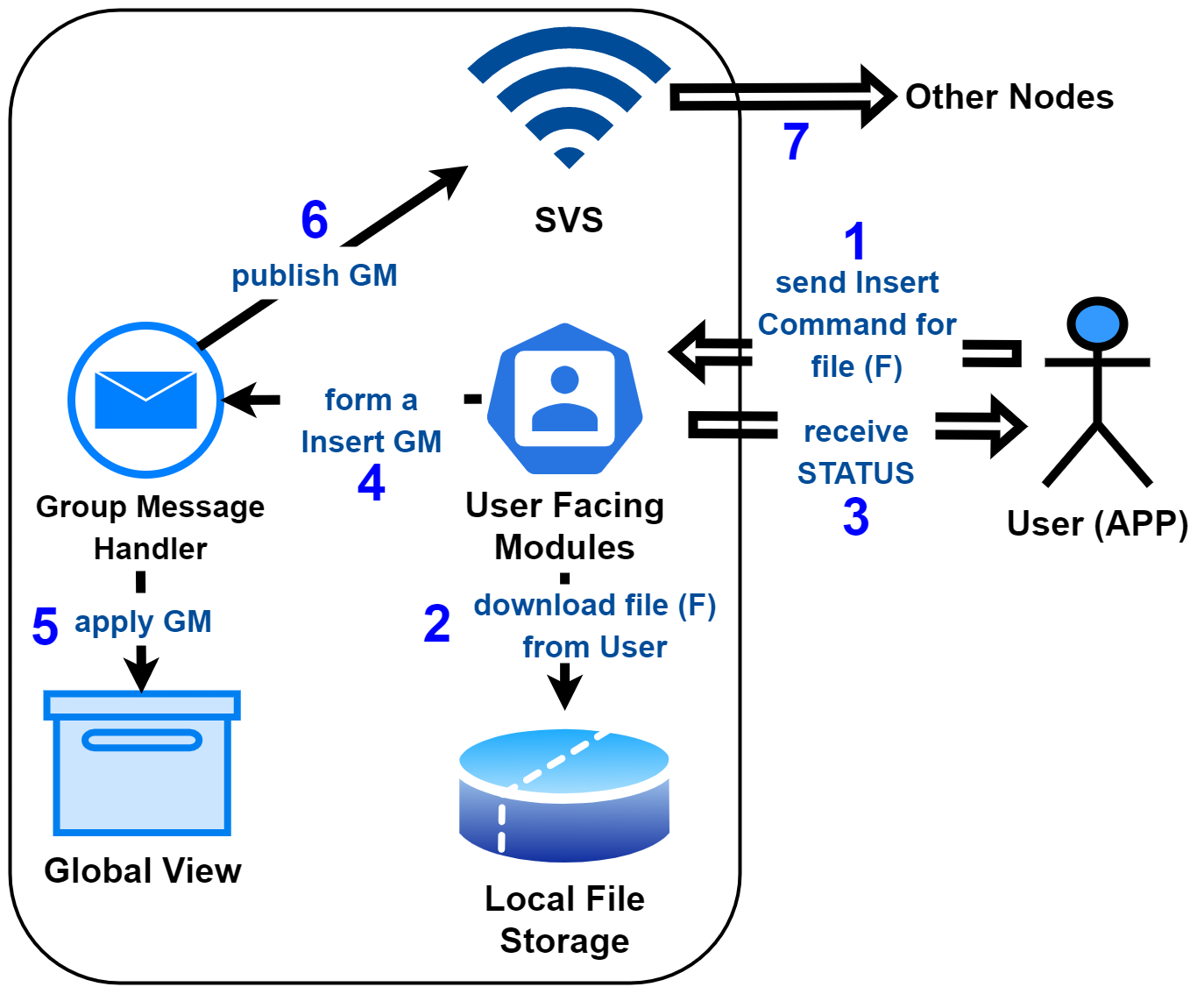}
    \caption{Module Interaction to Fulfill a User's Insertion Command}
    \label{fig:insert-sys}
\end{figure}

The User-to-Node interaction leads to several interactions within Hydra.
Figure~\ref{fig:insert-sys} shows these interactions. When node (X) receives an Insertion command for file (F) from user (A):
\begin{enumerate}
    \item Node (X) properly authenticates the command, checks to see if the command can be executed, and then immediately starts fetching file (F) if the command satisfies all requirements.
    \item After storing file (F), node (X) updates the status for user (A) allowing user (A) to go offline.
    \item Node (X) then forms an Insert GM which includes all file metadata such as file name, size, etc. and states that node (X) will be storing this file.
    \item Node (X) applies this GM to its Global View.
    \item Node (X) publishes this GM using SVS, our distributed synchronization protocol.
\end{enumerate}

Every node will receive the Insert GM. When node (Y) receives the Insert GM sent by node (X) containing file (F)'s info, it performs the following operations:
\begin{enumerate}
    \item Node (Y) applies the GM to its Global View: file (F) information is added.
    \item Node (Y) sees that there is a replication need as file (F) does not meet the replication degree of 3 (stated in Hydra's base policy). For how replication is done, see Subsection~\ref{sec:data-replicate}
\end{enumerate}

\subsubsection{Data Structure Formats}
There are several structures that are used for the entire data insertion process. For the group message structures used in Module interaction, please refer back to Section~\ref{sec:group-messages}.

The structures used in User-to-Node interaction include the following:
\begin{enumerate}
    \item FirstContact: This includes the preferred name prefix that the sender wants to use for further interaction and a sender command unique resource identifier (curi) that the sender will use to refer to this interaction.
    \item InsertCommand: This include all necessary info that a Hydra node needs for a publication which includes a fetch path, file name, and other metainfo about the file.
    \item NotificationSpecification: This simply includes a command unique resource identifier (curi) of the receiver that the sender is referring to.
    \item CommandStatus: This just contains a Status code that can give simple feedback to the user. Theses codes are very simple and based on HTTP response status codes, please see Table \ref{tab:status-codes}. 
\end{enumerate}
\subsection{Data Replication} \label{sec:data-replicate}

\subsubsection{Scenario}
The FABRIC platform that the Hydra instance is running on is seeing a dramatic increase in popularity. As such, new sites have been installed and more users have been added. With tens of sites and users, site failures can happen.
Hydra needs to have a fast replication mechanism in order to quickly recover from Hydra node failures.
Hydra should be able to copy Alice's file that she inserted in Subsection~\ref{sec:data-insert} several times (3 is Hydra's base policy) to prevent permanent data loss. To satisfy these scenarios, the following interactions are conducted.

\subsubsection{Module Interaction}
A replication need is noticed by the Global View Checker (see Figure~\ref{fig:checker-sys} for more details).

This need is a result of one of the following: 
\begin{enumerate}
    \item An Insert GM is heard.
    \item A Node has become unresponsive.
\end{enumerate}


When node (X) sees replication needs, it creates a list of the files that:
\begin{enumerate}
    \item Do not meet the necessary degree of replication counting nodes that are currently fetching the file and not counting unresponsive nodes
    \item Calculates Favor of each node based on the parameters it received via sync messages.
    \item If it is among the highest favor at this time, it starts the replication.
    
\end{enumerate}

After deciding on replication, the node does the following:
\begin{enumerate}
    \item Node (x) decides to replicate the file. 
    \item Node (X) starts to fetch the selected files by sending interests following the format /<node-name>/<hydra-prefix>/fetch/<file-name>/<segment-no> where the node name is a 
    selected node that has the file already.
    \item Node (X) updates its global view and  publishes a GM using SVS
\end{enumerate}



\subsection{Data Retrieval} \label{sec:data-retrieve}

\subsubsection{Scenario}
Bob in Dallas has heard people talk about the research Alice from Subsection~\ref{sec:data-insert} has been doing on the Hydra instance on FABRIC. So much so that Bob wants to see for himself what the data looks like. Of course, Bob (being a new user of the Hydra instance) has no idea where the data replicas are located. The replicas could not be on Dallas, the Hydra node he is most close to. Therefore, Bob needs a way to fetch Alice's data regardless of where the data is located within Hydra. To satisfy this scenario, the following interactions are conducted.

\subsubsection{User-to-Node Interaction}
The process of how a user fetches a file from Hydra can be described as a series of interests and data packets with the only difference being the segment number component. An important note is that any Hydra node can handle a user's fetch requests regardless of whether the contacted node has the file or not.

The user starts out by sending an interest who's name follows the form 
\name{/<file-name>/<segment-no>} with the filename being the name found within Hydra and the segment number starting at 0. The user will automatically assume the file spans multiple packets and always add the segment component to its interest. If the file only spans 1 packet, the user will find that out via the final block id within the first data packet. While fetching any a file starts the same, it does not end the same: there are 3 situations that can occur after expressing that interest.

The user's interest for file (F) can be fulfilled from the following:
\begin{enumerate}
    \item NACK: File (F) does not exist on Hydra.
    \item BLOB: File (F) exists on the contacted node, data is returned.
    \item ForwardingHint: File (F) exists in Hydra, but not on the contacted node. This acts as a redirect.
\end{enumerate}

To minimize traffic and be more precise with data retrieval, the user can first send an Interest and see how the Interest will be fulfilled which can tell the user how to proceed. After finding out the correct fetch path either by getting data or by a redirect, a user can send Interests to get the rest of the file. If a NACK is received after sending the first Interest, the user knows that the file is not within Hydra and to stop further interactions.

\subsubsection{Module Interaction}
\begin{figure}[!ht]
    \centering
    \includegraphics[width=\columnwidth]{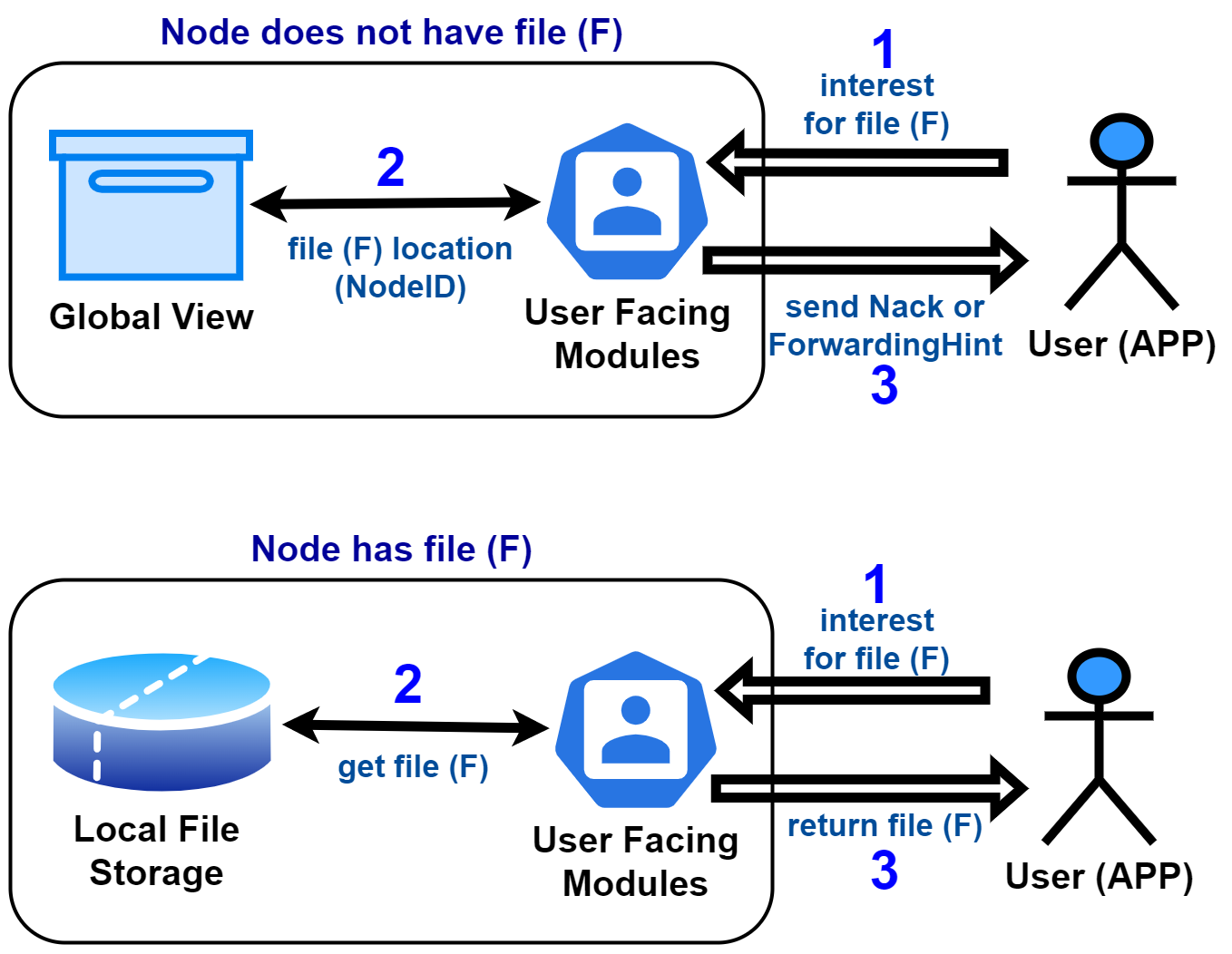}
    \caption{Module Interaction to Fulfill a User's Retrieval Requests}
    \label{fig:fetch-sys}
\end{figure}

Figure~\ref{fig:fetch-sys} shows two different situations. The Global View is used to indicate where the requested file is. After finding this information, the Hydra node responds to the user in the already-list 3 ways. In the case that the contacted node has the file, the local file storage is used to provide the file data. If a ForwardingHint is required, the node selects a random Hydra node that has the file and provides a forwarding hint (using the selected node's name) following the format \name{/<hydra-prefix>/<node-name>} for the user to use to fetch the file. Please note that this is the same format that a node uses to fetch a file from another node.
\subsection{Data Querying} \label{sec:data-query}

\subsubsection{Scenario}
It has been two weeks since Alice from Subsection~\ref{sec:data-insert} published the processed (i.e. indexed) axolotl genome. In fact, she completely forgot that she did insert this data into the Hydra instance on FABRIC. When Alice received an error for trying to insert new data with the same name, she remembered about her forgotten inserted file. From Alice's perspective, having some way of querying to see what files are in Hydra and what file metadata is under a certain name is extremely useful. To satisfy this scenario, the following interactions are conducted.

\subsubsection{User-to-Node Interaction}
The process of how a user sends a query to a Hydra node can be described as a single Interest and a corresponding data packet. Any Hydra node can handle a query sent by the user. The naming of a query interest follows the format \name{/<hydra-prefix>/query/<query-type>}.

\subsubsection{Module Interaction}
\begin{figure}[!ht]
    \centering
    \includegraphics[width=\columnwidth]{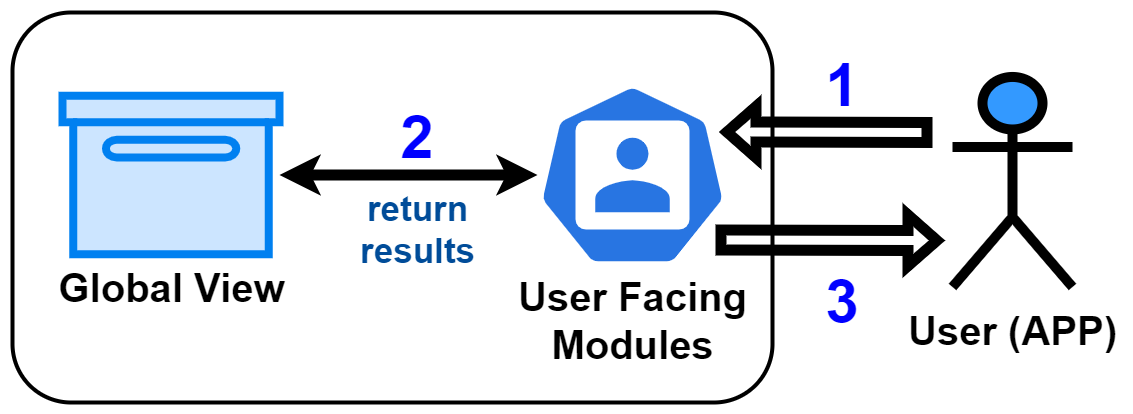}
    \caption{Module Interaction to Fulfill a User's Query Request}
    \label{fig:query-sys}
\end{figure}

The Interest by the user gets processed in a simple way which is described by Figure~\ref{fig:query-sys}. The Hydra node simply checks the Global View for the query type information and returns it via a data packet.

\subsubsection{Types}
There are different query types that are defined. It is important to note that more queries can be added to better suit the environment, but also any queries can be disabled via Hydra's base policy. This gives Hydra instances more flexibility in what they want to expose to the users.

The query types are the following:
\begin{enumerate}
    \item /files: This allows users to see what files are within a Hydra instance. 
    \item /nodes: This allows operators to see what nodes are part of a Hydra instance. 
    \item /prefix/<prefix>: This allows users to search Hydra for files under a certain prefix.
    \item /file/<filename>: This allows users to see information about a certain file.
\end{enumerate}
\subsection{Data Deletion} \label{sec:data-delete}

\subsubsection{Scenario} 
The same Alice found in Subsection~\ref{sec:data-insert} finds out that she made a grave mistake in pre-processing (i.e. indexing) the axolotl genome three weeks after she had inserted the data into Hydra. While Alice can wait a few more weeks to let the data naturally remove itself due to Hydra's base policy, she fears that anyone using her data before it gets removed will get false hope and possibly base future research on false data. This is a major concern for Clemson's Genomics and Bioinformatics facility as it can harmfully affect the facility's goal with using Hydra. Therefore, Alice needs a way to delete a file. To satisfy this scenario, the following interactions are conducted.

\subsubsection{User-to-Node Interaction} 
\begin{figure}[!ht]
    \centering
    \includegraphics[width=\columnwidth]{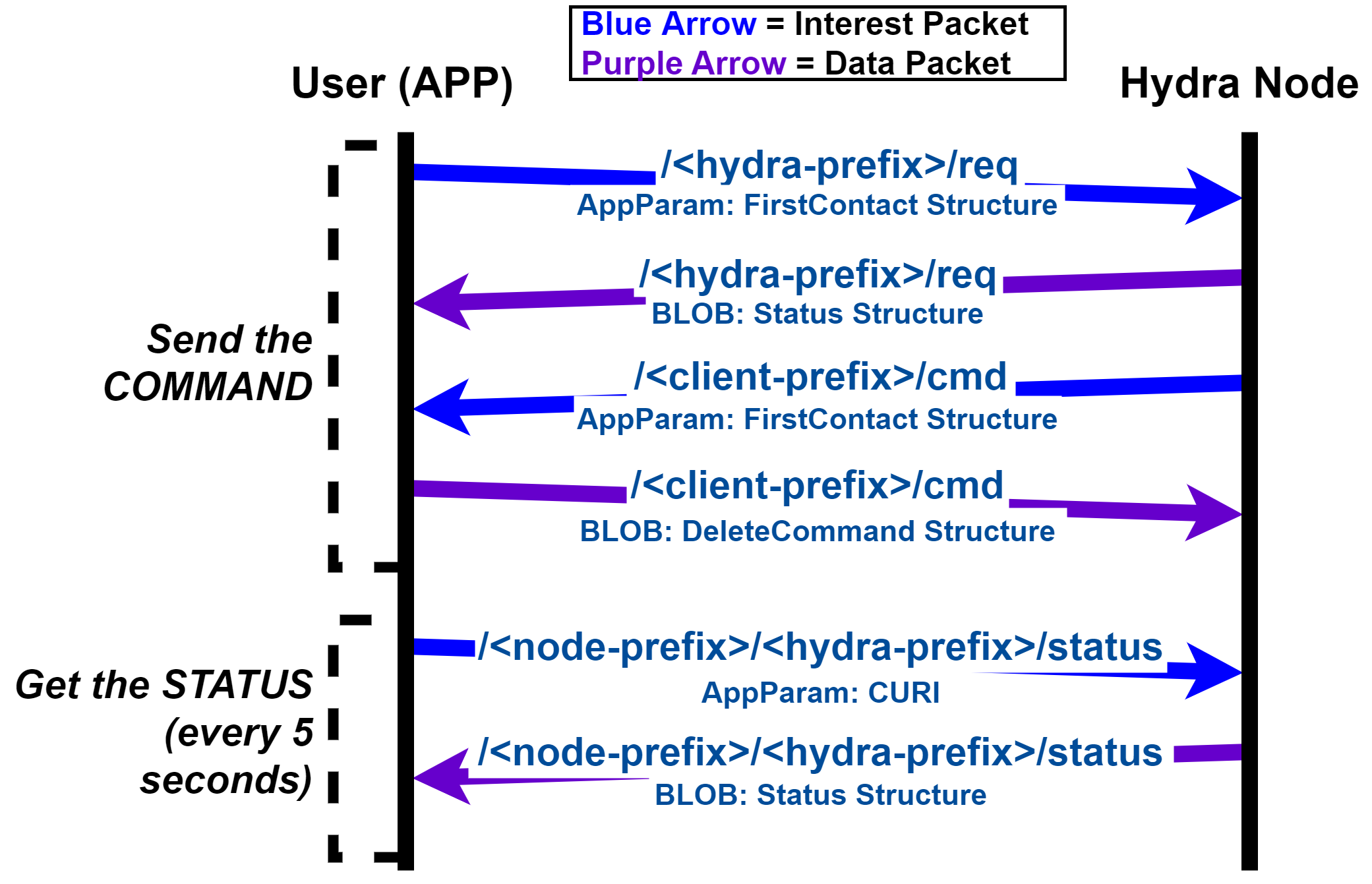}
    \caption{NDN Interaction of a User and a Hydra Node During Deletion}
    \label{fig:delete-usr}
\end{figure}

The process of how a user will interact with a Hydra node via NDN is described in Figure~\ref{fig:delete-usr} and shares great similarities with data insertion. Any Hydra node can be contacted by the user for data deletion, and this interaction can be summarized as a PubSub-like interaction: it uses a notification interest with a component /notify and a data retrieval interest with a component /msg.

This interaction between user (A) and node (X) goes as follows:
\begin{enumerate}
    \item User (A) will send a interest notification using the prefix /<hydra-prefix/delete, announcing that it has a command for any Hydra node to process.
    \item Upon hearing this, node (X) will send an interest to fetch this command using user (A)'s prefix (stated in user (A)'s FirstContact structure found in the interest notification's app parameters).
    \item Node (X) will process this command and than send a notification interest stating that the command that user (A) sent has a updated status.
    \item User (A) can then fetch the status of the command using node (X)'s prefix (stated in node (X)'s FirstContact structure found in a previous interest).
\end{enumerate}

The exact same key points found in Subsection~\ref{sec:data-insert} for User-to-Node interaction applies for this process as well.

\subsubsection{Module Interaction} 
\begin{figure}[!ht]
    \centering
    \includegraphics[width=\columnwidth]{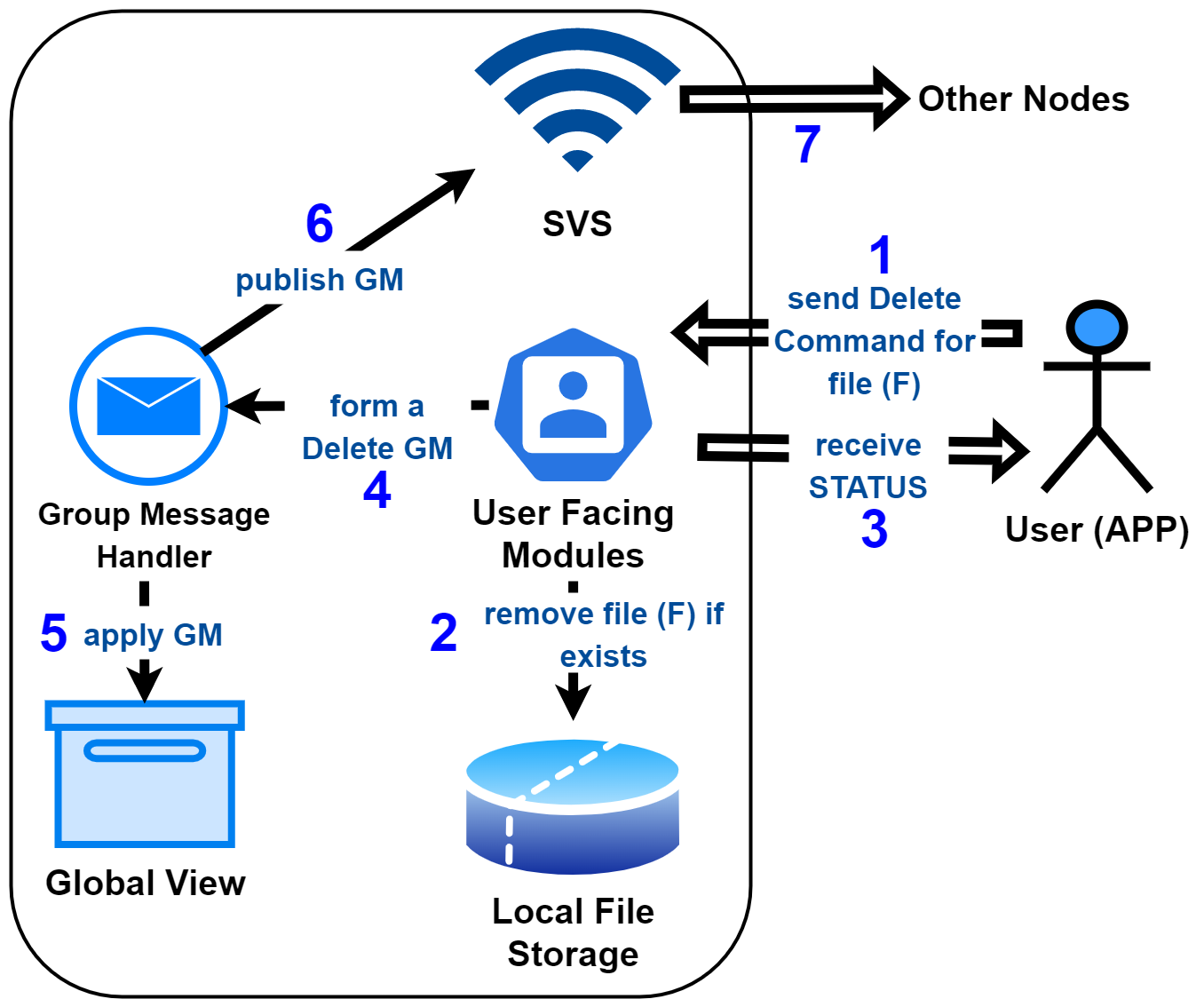}
    \caption{Module Interaction to Fulfill a User's Deletion Command}
    \label{fig:delete-sys}
\end{figure}

The User-to-Node interaction leads to Module interaction where a Hydra node will act out the given command. As seen in Figure~\ref{fig:delete-sys}, when node (X) receives a Deletion command for file (F) from user (A):
\begin{enumerate}
    \item Node (X) properly authenticates the command.
    \item Node (X) deletes any local copies of file (F).
    \item Node (X) updates the status for user (A) allowing user (A) to go offline.
    \item Node (X) forms a Delete GM which includes the name of file (F).
    \item Node (X) applies this GM to its Global View: all file (F) information is removed.
    \item Node (X) publishes this GM using SVS, our distributed synchronization protocol.
\end{enumerate}

Every node will receive the Delete GM. When node (Y) receives the Delete GM sent by node (X) containing file (F)'s name:
\begin{enumerate}
    \item Node (Y) deletes any local copies of file (F).
    \item Node (Y) applies the GM to its Global View: all file (F) information is removed.
\end{enumerate}

\subsubsection{Data Structure Formats} 
There are several structures that are used for the entire data deletion process. For the group message structures used in Module interaction, please refer back to Section~\ref{sec:group-messages}. In addition, most of the other structures are the same ones described in Subsection~\ref{sec:data-insert}. The only missing structure is DeleteCommand which holds the filename of the desired-to-be-removed file.

\section{Deployment on FABRIC} \label{sec:fabric}
This section describes how we deployed Hydra on the NSF FABRIC testbed\cite{8972790}.

\subsection{Hydra Deployment Overview}
\begin{figure}[!ht]
    \centering
    \includegraphics[width=\columnwidth]{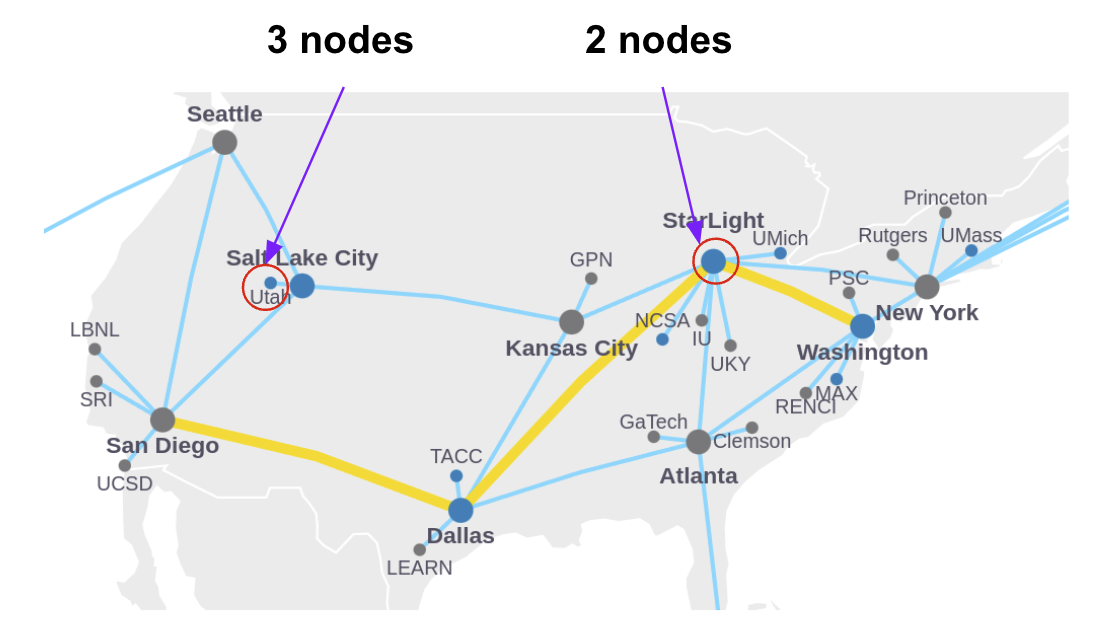}
    \caption{Hydra deployment on the FABRIC testbed.}
    \label{fig:nodes-on-fabric}
\end{figure}

As it shown in Figure \ref{fig:nodes-on-fabric}, We provisioned nodes on the NSF FABRIC (\url{https://fabric-testbed.net}) testbed as Hydra’s initial “soft” deployment. One node is a client node and others are hydra nodes. We choose five nodes because four nodes is the minimum number of nodes to demonstrate the Hydra function and one node as a client node. 

Our base FABRIC resource "slice" has the following properties:
\begin{itemize}
    \item Layer 2 connectivity (i.e. MAC address) to communicate, completely connected network graph, manual setup / routing.
    \item Easily deployed and destroyed to enable full system rebuilds for reproducibility testing. 
    \item High network bandwidth (100Gbps) links between nationally dispersed sites.
    \item Scalable resource allocation and modern hardware (e.g. NVME, GPUs, etc.) for the creation of a data lake of sufficient scale for data-intensive scientific workflows.
\end{itemize}

\subsection{Hydra Deployment on the FABRIC Testbed}
\begin{figure}[!ht]
    \centering
    \includegraphics[width=0.8\columnwidth]{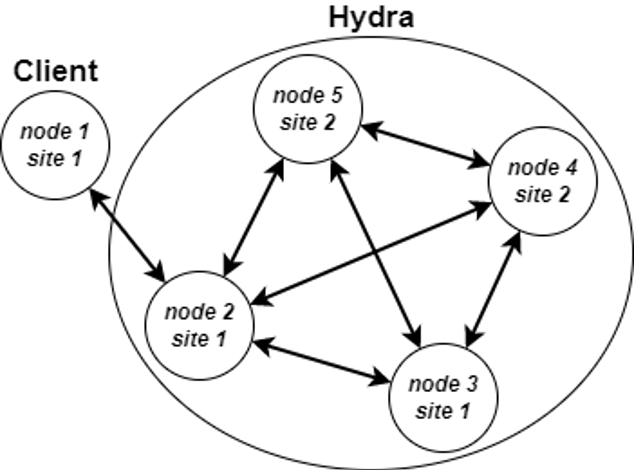}
    \caption{FABRIC topology overview of the client and Hydra nodes. }
    \label{fig:nodes-interface}
\end{figure}

To deploy the program on Fabric, first we have to become Fabric users by simply signing up at \url{http://portal.fabric-testbed.net}. The second step is to join a project. Users can join an existing project or they can create a new one. Fabric has its own JupterHub, which makes users easily integrate with the Fabric environment. The next step is setting up two SSH keypairs: bastion keypaire and sliver keypaire. The bastion keypair can be generated from the Fabric portal but has a finite lifetime. Users need to re-generate their bastion keypairs every 6 months. The sliver keypairs are long lived and installed into Fabric VMs. The environment configuration is completed when users add the path of these two keypairs into the Juyperhub. In our experiment, we used the FABRIC API (https://github.com/fabric-testbed/fabrictestbed-extensions) to request a slice with five nodes with three nodes in Utah and two nodes in Illinois. The five nodes were set up with default components: two CPU cores, 8Gb RAM, and 10Gb SSD disk space.  A layer 2 network was built between these five nodes as shown in Figure \ref{fig:nodes-interface}. 

After the deployment of the slice with basic VMs, we installed the Hydra packages on each node as python3-pip, libndn-cxx-dev, nfd, ndnping, ndnpeek, ndn-dissect, ndnchunks and ndnsec. Then, we established Hydra python libraries by running a command:

\begin{lstlisting}[language=bash]
    node:$ pip3 install python-ndn ndn-storage ndn-svs ndn-hydra
\end{lstlisting}

Next we added the NDN face and route between the client node and the Hydra nodes as in Figure \ref{fig:nodes-interface}. The following command shows how to create face and add route between two hydra nodes (node 2 and node 3) :
\begin{lstlisting}[language=bash]
  node2:$ nfd-start
  node2:$ nfdc face create remote ether://["MAC address of node 3"] local dev://"Ethernet interface of node 2 with node 3"
  node2:$ nfdc route add /hydra/group ether://["MAC address of node 3"]
  node2:$ nfdc route add /hydra/node/node3 ether://["MAC address of node 3"]
  node2:$ nfdc route add /node3 ether://["MAC address of node 3"]
\end{lstlisting}

\noindent The following command shows how to create face and add route between a hydra node and a client node (node 2 and node 1) :
\begin{lstlisting}[language=bash]
  node2:$ nfd-start
  node2:$ nfdc face create remote ether://["MAC address of node 1"] local dev://"Ethernet interface of node 2 with node 1"
  node2:$ nfdc route add /client ether://["MAC address of node 1"]
\end{lstlisting}

Finally we moved data into and out of the Hydra deployment using 1MB, 5MB and 10 MB files for testing purposes.

\subsection{Proof of Principle Hydra Experiment }
In our proof of principle experiment, we achieved four functionalities: \textit{query}, \textit{insert}, \textit{fetch}, and \textit{delete}. Once all the Hydra nodes came online, we were able to validate internode communication and performance with \textit{iperf}. 

First, the client node sent the \textit{query} to the Hydra node by command:
\begin{lstlisting}[language=bash]
  node1:$ ndn-hydra-client query -r REPONAME -q QUERY [-n NODENAME]
\end{lstlisting}

This command helps the client check whether the required files are stored on Hydra nodes. 

The second test command is the \textit{insert} command: 
\begin{lstlisting}[language=bash]
  node1:$ ndn-hydra-client insert -r REPONAME -f FILENAME -p PATH
\end{lstlisting}
The client node could then publish data to Hydra nodes. Note that the data file was separated into several small chunks based on the file sizes and stored on Hydra nodes.

Next, we \textit{fetch} the file from the Hydra node by using the command: 
\begin{lstlisting}[language=bash]
  node1:$ ndn-hydra-client fetch -r REPONAME -f FILENAME [-p PATH]
\end{lstlisting}
Through this command, Hydra linked the client node with the node with the request file and the request file was forwarded to the client node. 

After testing these three commands, the \textit{delete} command was used to remove all files on hydra nodes by using this command: 
\begin{lstlisting}[language=bash]
  node1:$ ndn-hydra-client delete -r REPONAME -f FILENAME
\end{lstlisting}

This basic procedure is being used to validate Hydra releases as well as vary testbed topology, slice hardware, dataset size, and data name schemas.

\subsection{Initial Evaluation}

In our proof of concept experiment, we achieved four functionalities: \textit{query}, \textit{insert}, \textit{fetch}, and \textit{delete}. Once all the Hydra nodes came online, we were able to validate inter-node communication and performance with Hydra. Table \ref{tab:hydra-insert}, it shows the transfer speed for inserting a file into Hydra nodes. We used 10 MB files, 100 MB files, and 1 GB files. For each file size, we ran the insert command three times and took the average value and standard deviation. We tested transfer speed on the client node side, on the first hydra node (rep1) storing the inserting file and on the second hydra node (rep2) replicating the inserting file.   

\begin{table}[]
    \centering
    \begin{tabular}{c c c c c c c}
     \toprule
     \multirow{2}{*}{file size} &
      \multicolumn{2}{c}{client} & 
      \multicolumn{2}{c}{rep1} &
      \multicolumn{2}{c}{rep2} \\
     
     & {avg} & {std} & {avg} & {std} & {avg} & {std}\\
     \midrule
     10 MB    & 12.50 & 2.26 &  27.28 & 0.53 & 19.52 & 7.41  \\
     100 MB &  16.86 & 0.22 & 135.65 & 6.52 & 61.72 & 61.85 \\
     1 GB &  11.29 & 0.01 & 62.17 & 60.58 & 27.26 & 0.24\\
     \bottomrule
    \end{tabular}
    \caption{Hydra insertion speed: We recorded the completion time of the insert command on the client node and the two hydra nodes (rep1 and rep2), and calculated the transfer speed (Mbps). We ran the insert command 3 times for each file size and took the average and standard deviation.}
    \label{tab:hydra-insert}
\end{table}


     


Table \ref{tab:hydra-fetch} shows the transfer speed for fetching a file from Hydra nodes. The file sizes are the same as the inserting file sizes. For each file size, we ran the fetch command three times and took the average value and standard deviation. For fetching process, we only want to know how fast the client node receives the request file, so we only consider the time consuming on the client node. For Hydra, we think the same file might serve different users. Therefore, when a client first fetches a file, the file is first fetched from disk to the cache and then from the cache to the client node. Once the first fetch command is finished, the file will stay in the cache. In the future fetch, if users request the same file, the file is fetched from the cache to the client node, which means the transfer speed will be faster than the first fetch. However, we set the cache size to 500 MB for these experiments. When the file size is greater than the cache size (e.g. 1 GB), there would not be a considerable difference between cached and uncached retrieval.


\begin{table}[]
    \centering
        \caption{Hydra fetching speed (Mbps): We recorded the completion time of the fetching command on the client node with and without caching.}
    \begin{tabular}{c c c c c c c}
     \toprule
     \multirow{2}{*}{File Size} &
     \multicolumn{2}{c}{No Caching(Mbps)} &
     \multicolumn{2}{c}{Caching(Mbps)} \\
     
     & {avg} & {std} & {avg} & {std}\\
     \midrule  
     10 MB & 13.23 & 0.29 & 183.92 & 2.11   \\
     100 MB &  13.85 & 0.11 & 185.82 & 5.12 \\
     1 GB &  13.37 & 0.30 & 13.51 & 0.21 \\ 
     \bottomrule
    \end{tabular}
    
    \label{tab:hydra-fetch}
\end{table}

\section{Conclusions and Future Work} \label{sec:conclusions}
There are several important future work we are working on. First, Favor needs to be explored further. We need to answer questions such as what should be the weights of various parameters, what is the efficiency and overhead of replication based on favor as compared to traditional methods?

Second, we continue to enhance the current data fetching module to reliably handle large files. 
Third, we are working on improving the replication mechanism to reduce replication latency while maintaining a desired degree of replication. 
Fourth, reduce the latency of SVS, maintain a more consistent global view to improve system efficiency. Fifth, in order to better maintain storage limits and further improve the performance of the garbage  collection process, a better estimate of the time limit for purging expired data is required. We are also working on defining the trust schemas for genomics use cases. Last but not least, we plan to improve Hydra's usability by addressing some of our assumptions. 


\bibliographystyle{plain}
\bibliography{references.bib}


\end{document}